\documentclass{amsart}
\usepackage{amsfonts}
\usepackage{amsmath}
\usepackage{amssymb}

\theoremstyle{plain}
\newtheorem{theorem}{Theorem}[section]

\theoremstyle{definition}

\theoremstyle{remark}
\newtheorem{remark}[theorem]{Remark}
\numberwithin{equation}{section}

\input{tcilatex}

\begin{document}
\title[The classical Maxwell electrodynamics]{The classical Maxwell-Lorentz
electrodynamics aspects of the electron inertia problem within the Feynman
proper time paradigm}
\author{Anatolij K. Prykarpatski}
\address{The Ivan Franko State Pedagogical University of Drohobych, Lviv
region, Ukraine, and the Faculty of Applied Mathematics at AGH University of
Science and Technology, Krakow 30059 Poland}
\email{ pryk.anat@ua.fm, prykanat@cybergal.com}
\author{Nikolai N. Bogolubov (Jr.)}
\address{The Abdus Salam International Centre for Theoretical Physics,
Trieste, Italy, and the V.A. Steklov Mathematical Institute of RAS, Moscow,
Russian Federation}
\email{nikolai\_bogolubov@hotmail.com}
\date{present}
\maketitle

\begin{abstract}
The Maxwell electromagnetic and the Lorentz type force equations are derived
in the framework of the R. Feynman proper time paradigm and the related
vacuum field theory approach. The electron inertia problem is analyzed
within the Lagrangian and Hamiltonian formalisms and the related
pressure-energy compensation principle. The modified Abraham-Lorentz damping
radiation force is derived, the electromagnetic elctron mass origin is
argued.
\end{abstract}

\section{Introduction}

The elementary point charged particle, like electron, mass problem was
inspiring many physicists \cite{Jamm} \ from the past as J. J. Thompson,
G.G. Stokes, H.A. Lorentz, E. Mach, M. Abraham, P.A. M. Dirac, G.A. Schott
and others. Nonetheless, their studies have not given rise to a clear
explanation of this phenomenon that stimulated new researchers to tackle it
from different approaches based on new ideas stemming both from the
classical Maxwell-Lorentz electromagnetic theory, as in \cite%
{Bril,GiZa-1,GiZa-2,Pryk-Ampe,Feyn-1,Feyn-2,Hamm-1,Hamm-2,Kosy-1,Kosy-2,MaPi,Medi,Moro,Page,Papp,Pegg,Puth,Simu,Teit,WhFe,YaTr}%
, and modern quantum field theories of Yang-Mills and Higgs type, as in \cite%
{Anni,Higg,Hoof,Wilc-2} and others, whose recent and extensive review is
done in \cite{Wilc-1}.

In the present work I will mostly concentrate on detail analysis and
consequences of the Feynman proper time paradigm \cite%
{Feyn-1,Feyn-2,Dyso-1,Dyso-2} subject to deriving the electromagnetic
Maxwell equations and the related Lorentz like force expression considered
from the vacuum field theory approach, developed in works \cite%
{BoPr-foun,BoPr-Feyn,BoPr-Lore,BoPr-math}, and further, on its applications
to the electromagnetic mass origin problem. Our treatment of this and
related problems, based on the least action principle within the Feynman
proper time paradigm \cite{Feyn-1}, has allowed to construct the
respectively modified Lorentz type equation for a moving in space and
radiating energy charged point particle. Our analysis also elucidates, in
particular, the computations of the self-interacting electron mass term in 
\cite{MaPi}, where there was proposed a not proper solution to the well
known classical Abraham-Lorentz \cite{Abra,Lore-1,Lore-2,Lore-3} and Dirac 
\cite{Dira} electron electromagnetic "4/3-electron mass" problem. As a
result of our scrutinized studying the classical electromagnetic mass
problem \ we have stated that it can be satisfactory solved within the
classical H. Lorentz and M. Abraham reasonings augmented with the additional
electron stability condition, which was not taken before into account yet
appeared to be very important for balancing the related electromagnetic
field and mechanical electron momenta. The latter, following recent enough
works \cite{Puth,Moro}, devoted to analyzing the electron charged shell
model, can be realized within there suggested \textit{pressure-energy
compensation principle, }suitably\textit{\ } \ applied to the \ ambient
electromagnetic energy fluctuations and the own electrostatic Coulomb
electron energy.

\section{Feynman proper time paradigm geometric analysis}

In this section, we will develop further the vacuum field theory approach
within the Feynman proper time paradigm, devised before in \cite%
{BoPr-Lore,BoPr-Feyn}, to the electromagnetic J.C. Maxwell and H. Lorentz
electron theories and show that they should be suitably modified: namely,
the basic Lorentz force equations should be generalized following the
Landau-Lifschitz least action recipe \cite{LaLi}, taking also into account
the pure electromagnetic field impact. When applied the devised vacuum field
theory approach to the classical electron shell model, the resulting Lorentz
force expression appears to \ \ satisfactorily explaine the electron
inertial mass term exactly coinciding with the electron relativistic mass,
thus confirming the well known assumption \cite{Jack,Rohr} by M. Abraham and
H. Lorentz.

As was reported by F. Dyson \cite{Dyso-1,Dyso-2}, the original Feynman
approach derivation of the electromagnetic Maxwell equations was based on an 
\textit{a priori} general form of the classical Newton type force, acting on
a charged point particle moving in three-dimensional \ space $\mathbb{R}^{3}$
endowed with the canonical Poisson brackets on the phase variables, \
defined on the associated tangent space $T\mathbb{(R}^{3}).$ As a result of
this approach there only the first part of the Maxwell equations were
derived, as the second part, owing to F. Dyson \cite{Dyso-1}, is related
with the charged matter nature, which appeared to be hidden. Trying to
complete this Feynman approach to the derivation of Maxwell's equations more
systematically we have observed \cite{BoPr-Feyn} that the original Feynman's
calculations, based on Poisson brackets analysis, were performed on the 
\textit{tangent space} $T\mathbb{(R}^{3})$ which is, subject to the problem
posed, not physically proper. The true Poisson brackets can be correctly
defined only on the \textit{coadjoint phase space} $T^{\ast }\mathbb{(R}%
^{3}),$ as seen from the classical Lagrangian equations and the related
Legendre transformation \cite{AbMa,Arno,Godb,BlPrSa} from $T\mathbb{(R}^{3})$
to $T^{\ast }\mathbb{(R}^{3}).$ Moreover, within this observation, the
corresponding dynamical Lorentz type equation for a charged point particle
should be written for the particle momentum, not for the particle velocity,
whose value is well defined only with respect to the proper relativistic
reference frame, associated with the charged point particle owing to the
fact that the Maxwell equations are Lorentz invariant.

Thus, from the very beginning, we shall reanalyze the structure of the
Lorentz force exerted on a moving charged point particle with a charge $\xi
\in\mathbb{R}$ by another point charged particle with a charge $\xi_{f}\in%
\mathbb{R}$, making use of the classical Lagrangian approach, and rederive
the corresponding electromagnetic Maxwell equations. The latter appears to
be strongly related to the charged point mass structure of the
electromagnetic origin as was suggested by R. Feynman and F. Dyson.

Consider a charged point particle moving in an electromagnetic field. For
its description, it is convenient to introduce a trivial fiber bundle
structure ${\pi }${$:\mathcal{M}\rightarrow \mathbb{R}^{3},\mathcal{M}=%
\mathbb{R}^{3}\times G$}$,$ {with the abelian structure group $G:=\mathbb{R}%
\backslash \{0${$\},$} equivariantly acting on the canonically symplectic
coadjoint space }$T^{\ast }(\mathcal{M})$ {\ endowed both with the canonical
symplectic structure }%
\begin{align}
\omega ^{(2)}(p,y;r,g)& :=d\text{ }pr^{\ast }\alpha ^{(1)}(r,g)=<dp,\wedge
dr>+  \label{2.0} \\
+& <dy,\wedge g^{-1}dg>_{\mathcal{G}}+<ydg^{-1},\wedge dg>_{\mathcal{G}} 
\notag
\end{align}%
{for all $(p,y;r,g)\in T^{\ast }(\mathcal{M}),$ \ where }$\alpha
^{(1)}(r,g):=<p,dr>+<y,g^{-1}dg>_{\mathcal{G}}\in T^{\ast }(\mathcal{M})$ is
the corresponding Liouville form on $\mathcal{M},${\ and with a connection
one-form }$\mathcal{A}:{M}\rightarrow {T^{\ast }(M)\times \mathcal{G}}\ \ $%
as 
\begin{equation}
{\mathcal{A}}(r,g):=g^{-1}<\xi A(r),dr>g+g^{-1}dg,  \label{2.1}
\end{equation}%
with $\xi \in \mathcal{G}^{\ast },(r,g)\in \mathbb{R}^{3}\times G,$ and $\ \
<\cdot ,\cdot >$ $\ $being the scalar product in $\mathbb{E}^{3}.$ The
corresponding curvature 2-form $\Sigma ^{(2)}\in \Lambda ^{2}(\mathbb{R}%
^{3})\otimes $ $\mathcal{G}$ \ is 
\begin{equation}
\Sigma ^{(2)}(r):=d{\mathcal{A}}(r,g)+{\mathcal{A}}(r,g)\wedge {\mathcal{A}}%
(r,g)=\xi \sum_{i,j=1}^{3}F_{ij}(r)dr^{i}\wedge dr^{j},  \label{2.2}
\end{equation}%
where%
\begin{equation}
F_{ij}(r):=\frac{\partial A_{j}}{\partial r^{i}}-\frac{\partial A_{i}}{%
\partial r^{j}}  \label{2.3}
\end{equation}%
{for $i,j=\overline{1,3}$ \ is the electromagnetic tensor with respect to
the reference frame }$\mathcal{K}_{t},$ characterized by the phase space {%
coordinates $(r,p)\in T^{\ast }(\mathbb{R}^{3})$. }As an element $\xi \in 
\mathcal{G}^{\ast }$ is still not fixed, it is natural to apply the standard 
\cite{AbMa,Arno,BlPrSa} invariant Marsden--Weinstein--Meyer reduction to the
orbit factor space $\ \ \tilde{P}_{\xi }:=P_{\xi }/G_{\xi }\ $ subject to
the related momentum mapping $l:T^{\ast }\mathbb{(}\mathcal{M})\rightarrow 
\mathcal{G}^{\ast },$ constructed with respect to the canonical symplectic
structure \ (\ref{2.0}) \ on $T^{\ast }\mathbb{(}\mathcal{M}),$ where, by
definition, $\xi \in \mathcal{G}^{\ast }$ is constant, $P_{\xi }:=l^{-1}(\xi
)\ \subset $ $T^{\ast }\mathbb{(}\mathcal{M})$ and $G_{\xi }=\{g\in
G:Ad_{G}^{\ast }\xi \}$ is the isotropy group of the element $\xi \in 
\mathcal{G}^{\ast }.$

As a result of the Marsden--Weinstein--Meyer reduction, one finds that $%
G_{\xi }\simeq G,$ the factor-space $\tilde{P}_{\xi }\simeq T^{\ast }\mathbb{%
(R}^{3})$ is endowed with a suitably reduced symplectic structure {$\bar{%
\omega}_{\xi }^{(2)}\in T^{\ast }($}$\tilde{P}_{\xi })$ and the
corresponding Poisson brackets on the reduced manifold $\tilde{P}_{\xi }\ $%
are 
\begin{align}
\{r^{i},r^{j}\}_{\xi }& =0,\text{ }\{p_{j},r^{i}\}_{\xi }=\delta _{j}^{i},
\label{2.4} \\
\{p_{i},p_{j}\}_{\xi }& =\xi F_{ij}(r)  \notag
\end{align}%
for $i,j=\overline{1,3},$ considered with respect to the reference frame $%
\mathcal{K}_{t}.$ {Introducing a new momentum variable }%
\begin{equation}
{\tilde{\pi}:=p+\xi A(r)}  \label{2.4a}
\end{equation}%
{\ on }$\tilde{P}_{\xi }${$,$} {it is easy to verify that \ $\bar{\omega}%
_{\xi }^{(2)}\rightarrow \tilde{\omega}_{\xi }^{(2)}:=<d{\tilde{\pi}},\wedge
dr>$}${,}${\ giving rise to the following \textit{\textquotedblleft minimal
interaction\textquotedblright } \ \ canonical Poisson brackets: \ 
\begin{equation}
\{r^{i},r^{j}\}_{{\tilde{\omega}_{\xi }^{(2)}}}=0,\;\{{\tilde{\pi}}%
_{j},r^{i}\}_{{\tilde{\omega}_{\xi }^{(2)}}}=\delta _{j}^{i},\;\{{\tilde{\pi}%
}_{i},{\tilde{\pi}}_{j}\}_{{\tilde{\omega}_{\xi }^{(2)}}}=0  \label{2.4b}
\end{equation}%
for $i,j=\overline{1,3}$ \ with respect to some new reference frame $%
\mathcal{\tilde{K}}_{t^{\prime }},$ characterized by the phase space
coordinates }${(r,\tilde{\pi})}\in \tilde{P}_{\xi }$ and a new evolution
parameter $t^{\prime }\in \mathbb{R}${\ \ if and only if the Maxwell field
compatibility equations 
\begin{equation}
\partial F_{ij}/\partial r_{k}+\partial F_{jk}/\partial r_{i}+\partial
F_{ki}/\partial r_{j}=0  \label{2.5}
\end{equation}%
are satisfied on }$\mathbb{R}^{3}${\ for all $i,j,k=\overline{1,3}$ with the
curvature tensor \ (\ref{2.3}).}

Now we proceed to a dynamic description of the interaction between two
moving charged point particles $\xi $ and $\xi _{f},$ moving respectively,
with the velocities $u:=dr/dt$ and $u_{f}:=dr_{f}/dt$ subject to the
reference frame $\mathcal{K}_{t}.$ Unfortunately, there is a fundamental
problem in correctly formulating a physically suitable action functional and
the related least \ action condition. There are clearly possibilities such as%
\begin{equation}
S_{p}^{(t)}:=\int_{t_{1}}^{t_{2}}dt\mathcal{L}_{p}^{(t)}[r;dr/dt]
\label{2.6a}
\end{equation}%
on a temporal interval $[t_{1},t_{2}]\subset \mathbb{R}$ with respect to the
laboratory reference frame $\mathcal{K}_{t},$%
\begin{equation}
S_{p}^{(t^{\prime })}:=\int_{t_{1}^{\prime }}^{t_{2}^{\prime }}dt^{\prime }%
\mathcal{L}_{p}^{(t^{\prime })}[r;dr/dt^{\prime }]  \label{2.6b}
\end{equation}%
on a temporal interval $[t_{1}^{\prime },t_{2}^{\prime }]\subset \mathbb{R}$
with respect to the moving reference frame $\mathcal{K}_{t^{\prime }}$ and 
\begin{equation}
S_{p}^{(\tau )}:=\int_{\tau _{1}}^{\tau _{2}}d\tau \mathcal{L}_{p}^{(\tau
)}[r;dr/d\tau ]  \label{2.6c}
\end{equation}%
on a temporal interval $[\tau _{1},\tau _{2}]\subset \mathbb{R}$ with
respect to the proper time reference frame $\mathcal{K}_{\tau },$ naturally
related to the moving charged point particle $\xi .$

It was first observed by Poincar\'{e} and Minkowski \- \cite{Paul} that the
temporal differential $\ d\tau $ is not a closed differential one-form,
which physically means that a particle can traverse many different paths in
space $\mathbb{R}^{3}$ with respect to the reference frame $\mathcal{K}_{t}$
during any given proper time interval $d\tau ,$ \textit{naturally} related
to its motion. This fact was stressed \cite{Eins-1,Eins-2,Mink,Paul,Poin} by
Einstein, Minkowski and Poincar\'{e}, and later exhaustively analyzed by R.
Feynman, who argued \cite{Feyn-1} that the dynamical equation of a moving
point charged particle is physically sensible only with respect to its
proper time reference frame. This is Feynman's proper time reference frame
paradigm, which was recently further elaborated and applied both to the
electromagnetic Maxwell equations in \cite{GiZa-1,GiZa-2,GiZaLi} and to the
Lorentz type equation for a moving charged point particle under external
electromagnetic field in \cite{BoPr-Feyn,BoPr-Lore,BoPr-foun,BlPrSa}. As it
was there argued from a physical point of view, the least action principle
should be applied only to the expression \ (\ref{2.6c}) written with respect
to the proper time reference frame $\mathcal{K}_{\tau },$ whose temporal
parameter $\tau \in \mathbb{R}$ is independent of an observer and is a
closed differential one-form. Consequently, this action functional is also
mathematically sensible, which in part reflects the Poincar\'{e}'s and
Minkowski's observation that the infinitesimal quadratic interval%
\begin{equation}
d\tau ^{2}=(dt^{\prime })^{2}-|dr-dr_{f}|^{2},  \label{2.6d}
\end{equation}%
relating the reference frames $\mathcal{K}_{t^{\prime }}$ and $\mathcal{K}%
_{\tau },$ can be invariantly used for the four-dimensional relativistic
geometry. The most natural way to contend with this problem is to first
consider the quasi-relativistic dynamics of the charged point particle $\xi $
\ with respect to the moving reference frame $\mathcal{K}_{t^{\prime }}$
subject to which the charged point particle $\xi _{f}$ \ is at rest.
Therefore, it possible to write down a suitable action functional \ (\ref%
{2.6b}), up to $O(1/c^{4}),$ as the light velocity $c\rightarrow \infty $,
where the quasi-classical Lagrangian function $\mathcal{L}_{p}^{(t^{\prime
})}[r;dr/dt^{\prime }]$ can be naturally chosen as 
\begin{equation}
\mathcal{L}_{p}^{(t^{\prime })}[r;dr/dt^{\prime }]:=m^{\prime }(r)\left\vert
dr/dt^{\prime }-dr_{f}/dt^{\prime }\right\vert ^{2}/2-\xi \varphi ^{\prime
}(r).  \label{2.8}
\end{equation}%
where $m^{\prime }(r)\in \mathbb{R}_{+}$ is the charged particle $\xi $
ineryial mass parameter and $\varphi ^{\prime }(r)$ is the potential
function generated by the charged particle $\xi _{f}$ \ at a point $r\in $ $%
\mathbb{R}^{3}$ with respect to the reference frame $\mathcal{K}_{t^{\prime
}}.$ Since the standard temporal relationships between reference frames $%
\mathcal{K}_{t}$ and $\mathcal{K}_{t^{\prime }}:$%
\begin{equation}
dt^{\prime }=dt(1-\left\vert dr_{f}/dt^{\prime }\right\vert ^{2})^{1/2},
\label{2.9}
\end{equation}%
as well as between the reference frames $\mathcal{K}_{t^{\prime }}$ and $%
\mathcal{K}_{\tau }:$ 
\begin{equation}
d\tau =dt^{\prime }(1-\left\vert dr/dt^{\prime }-dr_{f}/dt^{\prime
}\right\vert ^{2})^{1/2},  \label{2.10}
\end{equation}%
give rise, up to $O(1/c^{2}),$ as $c\rightarrow \infty ,$ to $dt^{\prime
}\simeq dt$ and $d\tau \simeq dt^{\prime },$ respectively, it is easy to
verify that the least action condition $\delta S_{p}^{(t^{\prime })}=0$ is
equivalent to the dynamical equation%
\begin{equation}
d\pi /dt=\nabla \mathcal{L}_{p}^{(t^{\prime })}[r;dr/dt]=(\frac{1}{2}%
\left\vert dr/dt-dr_{f}/dt\right\vert ^{2})\nabla m-\xi \nabla \varphi (r),
\label{2.11}
\end{equation}%
where we have defined the generalized canonical momentum as 
\begin{equation}
\pi :=\partial \mathcal{L}_{p}^{(t^{\prime })}[r;dr/dt]/\partial
(dr/dt)=m(dr/dt-dr_{f}/dt),  \label{2.12}
\end{equation}%
with the dash signs dropped and denoted by \textquotedblleft $\nabla $%
\textquotedblright\ the usual gradient operator in $\mathbb{E}^{3}.$
Equating the canonical momentum expression \ (\ref{2.12}) with respect to
the reference frame $\mathcal{K}_{t^{\prime }}$ to that of (\ref{2.4a}) \ {%
with respect to the canonical reference frame $\mathcal{\tilde{K}}%
_{t^{\prime }},$ \ and identifying the reference frame $\mathcal{\tilde{K}}$}%
$_{{t}^{\prime }}${\ with }$\mathcal{K}_{t^{\prime }},$ one obtains that 
\begin{equation}
m(dr/dt-dr_{f}/dt)=mdr/dt-\xi A(r),  \label{2.12a}
\end{equation}%
giving rise to the important inertial particle mass determining expression%
\begin{equation}
m\ =-\xi \varphi (r),\   \label{2.13}
\end{equation}%
which right away follows from the relationship%
\begin{equation}
\varphi (r)dr_{f}/dt=A(r).\   \label{2.14}
\end{equation}%
The latter is well known in the classical electromagnetic theory \cite%
{Jack,LaLi} for potentials $(\varphi ,A)\in T^{\ast }(M^{4})$ satisfying the
Lorentz condition%
\begin{equation}
\partial \varphi (r)/\partial t+<\nabla ,A(r)>=0,  \label{2.15}
\end{equation}%
yet the expression (\ref{2.13}) looks very nontrivial in relating the 
\textit{\textquotedblleft inertial\textquotedblright } \ mass of the charged
point particle $\xi $ \ to the electric potential, being both generated by
the ambient charged point particles $\xi _{f}.$ \ As was argued in articles 
\cite{BoPr-foun,BoPr-Feyn,PrBoTa-acti}, the above mass phenomenon is closely
related and from a physical perspective shows its deep relationship to the
classical electromagnetic mass problem.

Before further analysis of the completely relativistic the charge $\xi $
motion under consideration, we substitute the mass expression (\ref{2.13})
into the quasi-relativistic action functional (\ref{2.6b}) with the
Lagrangian (\ref{2.8}). As a result, we obtain two possible action
functional expressions, taking into account two main temporal parameters
choices:%
\begin{equation}
S_{p}^{(t^{\prime })}=-\int_{t_{1}^{\prime }}^{t_{2}^{\prime }}\xi \varphi
^{\prime }(r)(1+\frac{1}{2}\left\vert dr/dt^{\prime }-dr_{f}/dt^{\prime
}\right\vert ^{2})dt^{\prime }  \label{2.15a}
\end{equation}%
on an interval $[t_{1}^{\prime },t_{2}^{\prime }]\subset \mathbb{R},$ or 
\begin{equation}
S_{p}^{(\tau )}=-\int_{\tau _{1}}^{\tau _{2}}\xi \varphi ^{\prime }(r)(1+%
\frac{1}{2}\left\vert dr/d\tau -dr_{f}/d\tau \right\vert ^{2})d\tau
\label{2.15b}
\end{equation}%
on an $[\tau _{1},\tau _{2}]\subset \mathbb{R}$. \ The direct relativistic
transformations of (\ref{2.15b}) entail that 
\begin{align}
S_{p}^{(\tau )}& =-\int_{\tau _{1}}^{\tau _{2}}\xi \varphi ^{\prime }(r)(1+%
\frac{1}{2}\left\vert dr/d\tau -dr_{f}/d\tau \right\vert ^{2})d\tau \simeq
\label{2.15c} \\
& \simeq -\int_{\tau _{1}}^{\tau _{2}}\xi \varphi ^{\prime }(r)(1+\left\vert
dr/d\tau -dr_{f}/d\tau \right\vert ^{2})^{1/2}d\tau =  \notag \\
& =-\int_{\tau _{1}}^{\tau _{2}}\xi \varphi ^{\prime }(r)(1-|dr/dt^{\prime
}-dr_{f}/dt^{\prime }|)^{-1/2}d\tau =-\int_{t_{1}^{\prime }}^{t_{2}^{\prime
}}\xi \varphi ^{\prime }(r)dt^{\prime },  \notag
\end{align}%
giving rise to the correct, from the physical point of view, relativistic
action functional form (\ref{2.6b}), suitably transformed to the proper time
reference frame representation (\ref{2.6c}) via the Feynman proper time
paradigm. Thus, we have shown that the true action functional procedure
consists in a physically motivated choice of either the action functional
expression form (\ref{2.6a}) or (\ref{2.6b}). Then, it is transformed to the
proper time action functional representation form (\ref{2.6c}) within the
Feynman paradigm, and the least action principle is applied.

Concerning the above discussed problem of describing the motion of a charged
point particle $\xi $ \ in the electromagnetic field generated by another
moving charged point particle $\xi _{f},$ it must be mentioned that we have
chosen the quasi-relativistic functional expression (\ref{2.8}) in the form (%
\ref{2.6b}) with respect to the moving reference frame $\mathcal{K}%
_{t^{\prime }},$ because its form is physically reasonable and acceptable,
since the charged point particle $\xi _{f}$ \ is then at rest, generating no
magnetic field.

Based on the above relativistic action functional expression 
\begin{equation}
S_{p}^{(\tau )}:=-\int_{\tau _{1}}^{\tau _{2}}\xi \varphi ^{\prime
}(r)(1+\left\vert dr/d\tau -dr_{f}/d\tau \right\vert ^{2})^{1/2}d\tau
\label{2.15d}
\end{equation}%
written with respect to the proper reference from $\mathcal{K}_{\tau },$ one
finds the following evolution equation:%
\begin{equation}
d\pi _{p}/d\tau =-\xi \nabla \varphi ^{\prime }(r)(1+\left\vert dr/d\tau
-dr_{f}/d\tau \right\vert ^{2})^{1/2},  \label{2.15e}
\end{equation}%
where the generalized momentum is given exactly by the relationship \ (\ref%
{2.12}): 
\begin{equation}
\pi _{p}=m(dr/dt-dr_{f}/dt).  \label{2.16}
\end{equation}%
Making use of the relativistic transformation (\ref{2.9}) and the next one \
(\ref{2.10}), the equation (\ref{2.15e}) is easily transformed to 
\begin{equation}
\frac{d}{dt}(p+\xi A)=-\nabla \varphi (r)(1-\left\vert u_{f}\right\vert
^{2}),  \label{2.17}
\end{equation}%
where we took into account the related definitions: (\ref{2.13}) for the
charged particle $\xi $ mass, (\ref{2.14}) for the magnetic vector potential
and $\varphi (r)=$ $\varphi ^{\prime }(r)/(1-\left\vert u_{f}\right\vert
^{2})^{1/2}$ for the scalar electric potential with respect to the
laboratory reference frame $\mathcal{K}_{t}.$ Equation (\ref{2.17}) can be
further transformed, using elementary vector algebra, to the classical
Lorentz type form:%
\begin{equation}
dp/dt=\xi E+\xi u\times B-\xi \nabla <u-u_{f},A>,  \label{2.18}
\end{equation}%
where 
\begin{equation}
E:=-\partial A/\partial t-\nabla \varphi  \label{2.19}
\end{equation}%
is the related electric field and 
\begin{equation}
B:=\nabla \times A  \label{2.20}
\end{equation}%
is the related magnetic field, exerted by the moving charged point particle $%
\xi _{f}$ \ on the charged point particle $\xi $ \ with respect to the
laboratory reference frame $\mathcal{K}_{t}.$ The Lorentz type force
equation \ (\ref{2.18}) was obtained in \cite{BoPr-Feyn,BoPr-foun} in terms
of the moving reference frame $\mathcal{K}_{t^{\prime }},$ and recently
reanalyzed in \cite{BoPr-foun,Pryk-Ampe}. The obtained results follow in
part \cite{Rous-1,Rous-2} from Amp\`{e}re's classical works on constructing
the magnetic force between two neutral conductors with stationary currents.

\section{Analysis of the Maxwell and Lorentz force equations: the electron
inertial mass problem}

As a moving charged particle $\xi _{f}$ generates the suitable electric
field (\ref{2.19}) and magnetic field (\ref{2.20}) via their electromagnetic
potential $(\varphi ,A)\in T^{\ast }(M^{4})$ with respect to a laboratory
reference frame $\mathcal{K}_{t},$ we will supplement them naturally by
means of the external material equations describing the relativistic charge
conservation law:%
\begin{equation}
\partial \rho /\partial t+<\nabla ,J>=0,  \label{3.1}
\end{equation}%
where $(\rho ,J)\in T^{\ast }(M^{4})$ is a related four-vector for the
charge and current distribution in the space $\mathbb{R}^{3}.$ Moreover, one
can augment the equation (\ref{3.1}) with the experimentally well
established Gauss law%
\begin{equation}
<\nabla ,E>=\rho  \label{3.2}
\end{equation}%
to calculate the quantity $\Delta \varphi :=<\nabla ,\nabla \varphi >$ from
the expression (\ref{2.19}): 
\begin{equation}
\Delta \varphi =-\frac{\partial }{\partial t}<\nabla ,A>-<\nabla ,E>.
\label{3.3}
\end{equation}%
Having taken into account the relativistic Lorentz condition (\ref{2.15})
and the expression (\ref{3.2}) one easily finds that the wave equation 
\begin{equation}
\partial ^{2}\varphi /\partial t^{2}-\Delta \varphi =\rho  \label{3.4}
\end{equation}%
holds with respect to the laboratory reference frame $\mathcal{K}_{t}.$
Applying the rot-operation \textquotedblleft $\nabla \times $%
\textquotedblright\ to the expression (\ref{2.19}) we obtain, owing to the
expression (\ref{2.20}), the equation 
\begin{equation}
\nabla \times E+\partial B/\partial t=0,  \label{3.5}
\end{equation}%
giving rise, together with (\ref{3.2}), to the first pair of the classical
Maxwell equations. To obtain the second pair of the Maxwell equations, it is
first necessary to apply the rot-operation \textquotedblleft $\nabla \times $%
\textquotedblright to the expression (\ref{2.20}): 
\begin{equation}
\nabla \times B=\partial E/\partial t+(\partial ^{2}A/\partial t^{2}-\Delta
A)  \label{3.6}
\end{equation}%
and then apply $-\partial /\partial t$ to the wave equation (\ref{3.4}) to
obtain%
\begin{equation}
\begin{array}{c}
-\frac{\partial ^{2}}{\partial t^{2}}(\frac{\partial \varphi }{\partial t}%
)+<\nabla ,\nabla \frac{\partial \varphi }{\partial t}>=\frac{\partial ^{2}}{%
\partial t^{2}}<\nabla ,A>- \\ 
\\ 
-<\nabla ,\nabla <\nabla ,A>>=<\nabla ,\frac{\partial ^{2}A}{\partial t^{2}}%
-\nabla \times (\nabla \times A)-\Delta A>= \\ 
\\ 
=<\nabla ,\frac{\partial ^{2}A}{\partial t^{2}}-\Delta A>=<\nabla ,J>.%
\end{array}%
\   \label{3.7}
\end{equation}%
The result (\ref{3.7}) leads to the relationship 
\begin{equation}
\partial ^{2}A/\partial t^{2}-\Delta A=J,  \label{3.8}
\end{equation}%
if we take into account that both the vector potential $A\in \mathbb{E}^{3}$
and the vector of current $J\in \mathbb{E}^{3}$ are determined up to a
rot-vector expression $\nabla \times S$ for some smooth vector-function $%
S:M^{4}\rightarrow \mathbb{E}^{3}.$ Inserting the relationship \ (\ref{3.8})
into \ (\ref{3.6}), we obtain \ (\ref{3.5}) and the second pair of the
Maxwell equations:%
\begin{equation}
\nabla \times B=\partial E/\partial t+J,\text{ \ }\nabla \times E=\partial
B/\partial t.  \label{3.9}
\end{equation}%
It is important that the system of equations \ (\ref{3.9}) can be
represented by means of the least action principle $\delta S_{f-p}^{(t)}=0,$
where the action functional 
\begin{equation}
S_{f-p}^{(t)}:=\int_{t_{1}}^{t_{2}}dt\mathcal{L}_{f-p}^{(t)}  \label{3.10}
\end{equation}%
is defined on an interval $[t_{1},t_{2}]\subset \mathbb{R}$ \ by the
Landau-Lifschitz type \cite{LaLi} Lagrangian function 
\begin{equation}
\mathcal{L}_{f-p}^{(t)}=\int_{\mathbb{R}%
^{3}}d^{3}r((|E|^{2}-|B|^{2})/2+<J,A>-\rho \varphi )  \label{3.11}
\end{equation}%
with respect to the laboratory reference frame $\mathcal{K}_{t},$ which is
subject to the electromagnetic field a unique and physically reasonable.
From \ (\ref{3.11}) we deduce that the generalized field momentum 
\begin{equation}
\pi _{f}:=\partial \mathcal{L}_{f-p}^{(t)}/\partial (\partial A/\partial
t)=-E  \label{3.12}
\end{equation}%
and its evolution is given as 
\begin{equation}
\partial \pi _{f}/\partial t:=\delta \mathcal{L}_{f-p}^{(t)}/\delta A\
=J-\nabla \times B,  \label{3.13}
\end{equation}%
which is equivalent to the first Maxwell equation of \ (\ref{3.9}). As the
Maxwell equations allow the least action representation, it is easy to
derive\ \cite{AbMa,Arno,BlPrSa,BoPr-foun,PrBoTa-acti} their dual Hamiltonian
formulation with the Hamiltonian function 
\begin{equation}
H_{f-p}:=\int_{\mathbb{R}^{3}}d^{3}r<\pi _{f},\partial A/\partial t>-%
\mathcal{L}_{f-p}^{(t)}=\int_{\mathbb{R}%
^{3}}d^{3}r((|E|^{2}-|B|^{2})/2-<J,A>),  \label{3.14}
\end{equation}%
satisfying the invariant condition 
\begin{equation}
dH_{f-p}/dt=0  \label{3.15}
\end{equation}%
for all $t\in \mathbb{R}.$

It is worth noting here that the Maxwell equations were derived under the
important condition\ that the charged system $(\rho ,J)\in T(M^{4})$ exerts
no influence on the ambient electromagnetic field potentials $(\varphi
,A)\in T^{\ast }(M^{4}).$ As this is not actually the case owing to the
damping radiation reaction on accelerated charged particles, one can try to
describe this self-interacting influence by means of the modified least
action principle, making use of the Lagrangian expression \ (\ref{3.11})
recalculated with respect to the separately chosen charged particle $\xi $
endowed with the uniform shell model$\ $geometric structure and generating
this electromagnetic field.

Following the slightly modified well-known approach from \cite{LaLi} and
reasonings from \cite{Beck,Moro} this Landau-Lifschitz type Lagrangian \ (%
\ref{3.11}) can be recast (further in the Gauss units) as 
\begin{equation}
\begin{array}{c}
\mathcal{L}_{f-p}^{(t)}=\int_{\mathbb{R}^{3}}d^{3}r((|E|^{2}-|B|^{2})/2+%
\int_{\mathbb{R}^{3}}d^{3}r(\frac{1}{c}<J,A>-\rho \varphi )-<k(t),dr/dt>= \\ 
\\ 
=\int_{\mathbb{R}^{3}}d^{3}r(\frac{1}{2}<-\nabla \varphi -\frac{1}{c}%
\partial A/\partial t,-\nabla \varphi -\frac{1}{c}\partial A/\partial t>- \\ 
\\ 
-\frac{1}{2}<\nabla \times (\nabla \times A),A>)+\int_{\mathbb{R}^{3}}d^{3}r(%
\frac{1}{c}<J,A>-\rho \varphi )-<k(t),dr/dt>= \\ 
\\ 
=\int_{\mathbb{R}^{3}}d^{3}r(\frac{1}{2}<-\nabla \varphi ,E>-\frac{1}{2c}%
<\partial A/\partial t,E>-\frac{1}{2}<A,\nabla \times B>)+ \\ 
\\ 
+\int_{\mathbb{R}^{3}}(\frac{1}{c}<J,A>-\rho \varphi )-<k(t),dr/dt>=%
\end{array}
\label{3.16}
\end{equation}%
\begin{equation*}
\begin{array}{c}
=\int_{\mathbb{R}^{3}}d^{3}r(\frac{1}{2}\varphi <\nabla ,E>+\frac{1}{2c}%
<A,\partial E/\partial t>-\frac{1}{2c}<A,\ J+\partial E/\partial t>)+ \\ 
\\ 
+\int_{\mathbb{R}^{3}}(\frac{1}{c}<J,A>-\rho \varphi )-\frac{1}{2c}\frac{d}{%
dt}\int_{\mathbb{R}^{3}}d^{3}r<A,E>- \\ 
\\ 
-\frac{1}{2}\lim_{r\rightarrow \infty }\int_{\mathbb{S}_{r}^{2}}<\varphi
E+A\times B,dS_{r}^{2}>-<k(t),dr/dt>\ = \\ 
\\ 
=-\frac{1}{2}\int_{\mathbb{\Omega }_{+}(\xi )}d^{3}r(\frac{1}{c}<J,A>-\rho
\varphi )+\int_{\mathbb{\Omega }_{+}(\xi )\cup \Omega _{-}(\xi )}(\frac{1}{c}%
<J,A>-\rho \varphi )-<k(t),dr/dt>- \\ 
\\ 
-\frac{1}{2c}\frac{d}{dt}\int_{\mathbb{R}^{3}}d^{3}r<A,E>-\frac{1}{2}%
\lim_{r\rightarrow \infty }\int_{\mathbb{S}_{r}^{2}}<\varphi E+A\times
B,dS_{r}^{2}>=%
\end{array}%
\end{equation*}%
\begin{equation*}
\begin{array}{c}
=-\frac{1}{2}\int_{\mathbb{\Omega }_{+}(\xi )}d^{3}r(\frac{1}{c}<J,A>-\rho
\varphi )-\frac{1}{2}\int_{\mathbb{\Omega }_{-}(\xi )}d^{3}r(\frac{1}{c}%
<J,A>-\rho \varphi )+ \\ 
\\ 
+\frac{1}{2}\int_{\mathbb{\Omega }_{-}(\xi )}d^{3}r(\frac{1}{c}<J,A>-\rho
\varphi )+\int_{\mathbb{\Omega }_{+}(\xi )\cup \Omega _{-}(\xi )}(\frac{1}{c}%
<J,A>-\rho \varphi )-<k(t),dr/dt>- \\ 
\\ 
-\frac{1}{2c}\frac{d}{dt}\int_{\mathbb{R}^{3}}d^{3}r<A,E>-\frac{1}{2}%
\lim_{r\rightarrow \infty }\int_{\mathbb{S}_{r}^{2}}<\varphi E+A\times
B,dS_{r}^{2}>= \\ 
\\ 
=\frac{1}{2}\int_{\mathbb{\Omega }_{-}(\xi )}d^{3}r(\frac{1}{c}<J,A>-\rho
\varphi )-\frac{1}{2}\int_{\mathbb{\Omega }_{+}(\xi )\cup \Omega _{-}(\xi
)}d^{3}r(\frac{1}{c}<J,A>-\rho \varphi )+ \\ 
\\ 
+\int_{\mathbb{\Omega }_{+}(\xi )\cup \Omega _{-}(\xi )}(\frac{1}{c}%
<J,A>-\rho \varphi )-<k(t),dr/dt>- \\ 
\\ 
-\frac{1}{2c}\frac{d}{dt}\int_{\mathbb{R}^{3}}d^{3}r<A,E>-\frac{1}{2}%
\lim_{r\rightarrow \infty }\int_{\mathbb{S}_{r}^{2}}<\varphi E+A\times
B,dS_{r}^{2}>= \\ 
\\ 
=\frac{1}{2}\int_{\mathbb{\Omega }_{-}(\xi )}d^{3}r(\frac{1}{c}<J,A>-\rho
\varphi )+\frac{1}{2}\int_{\mathbb{\Omega }_{+}(\xi )\cup \Omega _{-}(\xi
)}d^{3}r(\frac{1}{c}<J,A>-\rho \varphi )- \\ 
\\ 
-\frac{1}{2c}\frac{d}{dt}\int_{\mathbb{R}^{3}}d^{3}r<A,E>-\frac{1}{2}%
\lim_{r\rightarrow \infty }\int_{\mathbb{S}_{r}^{2}}<\varphi E+A\times
B,dS_{r}^{2}>,%
\end{array}%
\end{equation*}%
where we have introduced still not determined a radiation damping force $%
k(t)\in \mathbb{E}^{3},$ have denoted by $\mathbb{\Omega }_{+}\mathbb{(\xi )}%
:=supp$ $\xi _{+}\subset \mathbb{R}^{3}\ $and $\ \mathbb{\Omega }_{-}\mathbb{%
(\xi )}:=supp$ $\xi _{-}$ $\subset \mathbb{R}^{3}$ the corresponding charge $%
\xi $ supports, located\ on the electromagnetic field shadowed rear and
electromagnetic field exerted front semispheres (see Fig.1) of the electron
shell, respectively to its motion with the fixed velocity $u(t)\in \mathbb{E}%
^{3},$ as well as we denoted by $\mathbb{S}_{r}^{2}$ \ a two-dimensional
sphere of radius $r\rightarrow \infty .$

[The courtesy picture from \cite{Moro}]

Having naturally assumed that the radiated charged particle energy at
infinity is negligible, the Lagrangian function \ (\ref{3.16}) becomes
equivalent to 
\begin{equation}
\begin{array}{c}
\mathcal{L}_{f-p}^{(t)}=\frac{1}{2}\int_{\mathbb{\Omega }_{-}(\xi )}d^{3}r(%
\frac{1}{c}<J,A>-\rho \varphi )+\frac{1}{2c}\int_{\mathbb{\Omega }_{+}(\xi
)\cup \Omega _{-}(\xi )}(<J,A>-\rho \varphi )-<k(t),dr/dt>,%
\end{array}
\label{3.17}
\end{equation}%
which we now need to additionally recalculate taking into account that the
electromagnetic potentials $(\varphi ,A)\in T^{\ast }(M^{4})$ are retarded,
generated by only the front part of the electron shell and given as $%
1/c^{2}\rightarrow 0$ in the following expanded into Lienard-Wiechert series
form:

\begin{equation}
\begin{array}{c}
\varphi =\left. \int_{\mathbb{R}^{3}}d^{3}r^{\prime }\frac{\rho (t^{\prime
},r^{\prime })}{|r-r^{\prime }|}\right\vert _{t^{\prime }=t-|r-r^{\prime
}|/c}=\lim_{\varepsilon \downarrow 0}\int_{\mathbb{R}^{3}}d^{3}r^{\prime }%
\frac{\rho (t-\varepsilon ,r^{\prime })}{|r-r^{\prime }|}+ \\ 
\\ 
+\lim_{\varepsilon \downarrow 0}\frac{1}{2c^{2}}\int_{\mathbb{R}%
^{3}}d^{3}r^{\prime }|r-r^{\prime }|\partial ^{2}\rho (t-\varepsilon
,r^{\prime })/\partial t^{2}+ \\ 
\\ 
+\lim_{\varepsilon \downarrow 0}\frac{1}{6c^{3}}\int_{\mathbb{R}%
^{3}}d^{3}r^{\prime }|r-r^{\prime }|^{2}\partial \rho (t-\varepsilon
,r^{\prime })/\partial t+O(1/c^{4})= \\ 
\\ 
=\int_{\mathbb{\Omega }_{+}(\xi )}d^{3}r^{\prime }\frac{\rho (t,r^{\prime })%
}{|r-r^{\prime }|}+\ \frac{1}{2c^{2}}\int_{\mathbb{\Omega }_{+}(\xi
)}d^{3}r^{\prime }|r-r^{\prime }|\partial ^{2}\rho (t,r^{\prime })/\partial
t^{2}+ \\ 
\\ 
+\frac{1}{6c^{3}}\int_{\mathbb{\Omega }_{+}(\xi )}d^{3}r^{\prime
}|r-r^{\prime }|^{2}\partial \rho (t,r^{\prime })/\partial t+O(1/c^{4}), \\ 
\end{array}
\label{3.18}
\end{equation}%
\begin{equation*}
\begin{array}{c}
A=\left. \frac{1}{c}\int_{\mathbb{R}^{3}}d^{3}r^{\prime }\frac{J(t^{\prime
},r^{\prime })}{|r-r^{\prime }|}\right\vert _{t^{\prime }=t-|r-r^{\prime
}|/c}=\lim_{\varepsilon \downarrow 0}\frac{1}{c}\int_{\mathbb{R}%
^{3}}d^{3}r^{\prime }\frac{J(t-\varepsilon ,r^{\prime })}{|r-r^{\prime }|}-
\\ 
\\ 
-\lim_{\varepsilon \downarrow 0}\frac{1}{c^{2}}\int_{\mathbb{R}%
^{3}}d^{3}r^{\prime }\partial J(t-\varepsilon ,r^{\prime })/\partial t+ \\ 
\\ 
+\lim_{\varepsilon \downarrow 0}\frac{1}{2c^{3}}\int_{\mathbb{R}%
^{3}}d^{3}r^{\prime }|r-r^{\prime }|\partial ^{2}J(t-\varepsilon ,r^{\prime
})/\partial t^{2}+O(1/c^{4})= \\ 
\\ 
=\frac{1}{c}\int_{\mathbb{\Omega }_{+}(\xi )}d^{3}r^{\prime }\frac{%
J(t,r^{\prime })}{|r-r^{\prime }|}-\frac{1}{c^{2}}\int_{\mathbb{\Omega }%
_{+}(\xi )}d^{3}r^{\prime }\partial J(t,r^{\prime })/\partial t+ \\ 
\\ 
+\frac{1}{2c^{3}}\int_{\mathbb{\Omega }_{+}(\xi )}d^{3}r^{\prime
}|r-r^{\prime }|\partial ^{2}J(t,r^{\prime })/\partial t^{2}+O(1/c^{4}),%
\end{array}%
\end{equation*}%
where \textit{\ }the current density $J(t,r)=\rho (t,r)dr/dt$ for all $t\in 
\mathbb{R}$ and $r\in \Omega (\xi ):=\mathbb{\Omega }_{+}(\xi )\cup \mathbb{%
\Omega }_{+}(\xi )\simeq \mathbb{S}^{2}:=supp$ $\rho (t;r)\subset \mathbb{R}%
^{3},$ being the spherical compact support of the charged particle density
distribution, and the limit $\lim_{\varepsilon \downarrow 0}$ \textit{was \
treated physically, that is taking into account the assumed shell modell of
the charged particle }$\xi $\textit{\ and its corresponding charge density
self interaction}. \ Moreover, the potentials \ (\ref{3.18}) are both
considered to be retarded and non singular, moving in space with the
velocity $u\in T(\mathbb{R}^{3})$ subject to the laboratory reference frame $%
\mathcal{K}_{t}.$ As a result of simple enough calculations like in \cite%
{Jack}, making use of \ the expressions (\ref{3.18}) one obtains that the
Lagranfian function (\ref{3.17}) brings about 
\begin{equation}
\mathcal{L}_{f-p}^{(t)}=\frac{\mathcal{E}_{es}}{2c^{2}}|u|^{2}-<k(t),dr/dt>,
\label{3.18b}
\end{equation}%
where we took into account that owing to the reasonings from \cite{Beck,Moro}
the only front half the electric charge interacts with the whole virtually
identical charge charge $\xi ,$ \ as well as made use of the following up to 
$O(1/c^{4})\ $limiting integral expressions: 
\begin{equation}
\ 
\begin{array}{c}
\ \int_{\mathbb{\Omega }_{+}(\xi )\cup \mathbb{\Omega }_{-}(\xi
)}d^{3}r\int_{\mathbb{\Omega }_{+}(\xi )\cup \mathbb{\Omega }_{-}(\xi )}\
d^{3}r^{\prime }\rho (t,r^{\prime })\rho (t,r^{\prime }):=\xi ^{2}, \\ 
\\ 
\mathcal{\ }\frac{1}{2}\int_{\mathbb{\Omega }_{+}(\xi )\cup \mathbb{\Omega }%
_{-}(\xi )}d^{3}r\int_{\mathbb{\Omega }_{+}(\xi )\cup \mathbb{\Omega }%
_{-}(\xi )}\ d^{3}r^{\prime }\frac{\rho (t,r^{\prime })\rho (t,r^{\prime })}{%
|r-r^{\prime }|}:=\mathcal{E}_{es}, \\ 
\\ 
\ \ \int_{\mathbb{\Omega }_{+}(\xi )}d^{3}r\rho (t,r)\int_{\mathbb{\Omega }%
_{+}(\xi )}d^{3}r^{\prime }\frac{\rho (t;r^{\prime })}{|r^{\prime }-r|}\ =%
\frac{1}{2}\mathcal{E}_{es}, \\ 
\\ 
\ \int_{\mathbb{\Omega }_{-}(\xi )}d^{3}r\rho (t,r)\int_{\mathbb{\Omega }%
_{-}(\xi )}d^{3}r^{\prime }\frac{\rho (t;r^{\prime })}{|r^{\prime }-r|}\ =%
\frac{1}{2}\mathcal{E}_{es}, \\ 
\\ 
\ \int_{\mathbb{\Omega }_{-}(\xi )}d^{3}r\rho (t,r)\ \int_{\mathbb{\Omega }%
_{+}(\xi )}d^{3}r^{\prime }\frac{\rho (t;r^{\prime })}{|r-r^{\prime }|}|%
\frac{<r^{\prime }-r,u>}{|r^{\prime }-r|}|^{2}>:=\frac{\mathcal{E}_{es}}{6}%
|u|^{2}, \\ 
\\ 
\ \int_{\mathbb{\Omega }_{+}(\xi )}d^{3}r\rho (t,r)\ \int_{\mathbb{\Omega }%
_{+}(\xi )}d^{3}r^{\prime }\frac{\rho (t;r^{\prime })}{|r-r^{\prime }|}|%
\frac{<r^{\prime }-r,u>}{|r^{\prime }-r|}|^{2}>:=\frac{\mathcal{E}_{es}}{6}%
|u|^{2}.%
\end{array}
\label{3.19}
\end{equation}

To obtain the corresponding evolution equation for our charged particle $\xi 
$ \ we need, within the Feynman proper time paradigm, to transform the
Lagrangian function \ (\ref{3.18b}) to the one with respect to the proper
time reference frame $\mathcal{K}_{\tau }:$%
\begin{equation}
\mathcal{L}_{f-p}^{(\tau )}=\ (m_{es}/2)|\dot{r}|^{2}(1+|\dot{r}%
|^{2}/c^{2})^{-1/2}-<k(t),\dot{r}>,  \label{3.20}
\end{equation}%
where, for brevity, we have denoted by $\dot{r}$ $:=dr/d\tau $ the charged
particle velocity with respect to the proper reference frame $\mathcal{K}%
_{\tau }$ and by, definition, $m_{es}:=\mathcal{E}_{es}/c^{2}$ its so called
electrostatic mass with respect to the laboratory refrence frame $\mathcal{K}%
_{t}.$

Thus, the generalized charged particle $\xi $ momentum \ (up to $O(1/c^{4}))$
equals 
\begin{equation}
\begin{array}{c}
\pi _{p}:=\partial \mathcal{L}_{f-p}^{(\tau )}/\partial \dot{r}=\ \frac{%
m_{es}\dot{r}}{(1+|\dot{r}|^{2}/c^{2})^{1/2}}-\frac{m_{es}|\dot{r}|^{2}\dot{r%
}}{2c^{2}(1+|\dot{r}|^{2}/c^{2})^{3/2}}-k(t)= \\ 
\\ 
=\ m_{es}u(1-\frac{|u|^{2}}{2c^{2}})-k(t)\simeq
m_{es}u(1-|u|^{2}/c^{2})^{1/2}-k(t)=\bar{m}_{es}u-k(t),%
\end{array}
\label{3.21}
\end{equation}%
where we denoted, as before, by $u:=dr/dt$ the charged particle $\xi $
velocity with respect to the laboratory reference frame $\mathcal{K}_{t}\ $%
and put, by definition, $\ $%
\begin{equation}
\bar{m}_{es}:=m_{es}(1-|u|^{2})^{1/2}\ \   \label{3.22}
\end{equation}%
its mass parameter $\bar{m}_{es}\in \mathbb{R}_{+}$ with respect to the
proper reference frame $\mathcal{K}_{\tau }.$

The generalized momentum \ (\ref{3.22}) satisfies with respect to the proper
reference frame $\mathcal{K}_{\tau }$ the evolution equation 
\begin{equation}
d\pi _{p}/d\tau :=\partial \mathcal{L}_{f-p}^{(\tau )}/\partial r=0,
\label{3.23}
\end{equation}%
$\ $being \ equivalent, with respect to the laboratory reference frame $%
\mathcal{K}_{t},\ $ \ to the Lorentz type equation 
\begin{equation}
\frac{d}{dt}(\bar{m}_{es}u)=\ dk(t)/dt.  \label{3.23a}
\end{equation}%
The evolution equation \ (\ref{3.23a}) allows the corresponding canonical
Hamiltonian formulation on the phase space $T^{\ast }(\mathbb{R}^{3})$ with
the Hamiltonian function 
\begin{equation}
\begin{array}{c}
H_{f-p}^{\ }:=<\pi _{p},r>-\mathcal{L}_{f-p}^{(\tau )}\simeq <\frac{m_{es}%
\dot{r}}{(1+|\dot{r}|^{2}/c^{2})^{1/2}}-\frac{m_{es}|\dot{r}|^{2}\dot{r}}{%
2c^{2}(1+|\dot{r}|^{2}/c^{2})^{3/2}}-k(t),\dot{r}>- \\ 
\\ 
-(m_{es}/2)|\dot{r}|^{2}(1+|\dot{r}|^{2}/c^{2})^{-1/2}+<k(t),\dot{r}>=\bar{m}%
_{es}|u|^{2}/2,%
\end{array}
\label{3.23b}
\end{equation}%
naturally looking and satisfying up to $O(1/c^{4})$ for all $\tau $ and $t$ $%
\in $ $\mathbb{R}$ the conservation conditions 
\begin{equation}
\frac{d}{d\tau }H_{f-p}=0=\frac{d}{dt}H_{f-p}.\ \   \label{3.23c}
\end{equation}%
Looking at the equation \ (\ref{3.23a}) and \ (\ref{3.23b}), one can state
that the physically observable inertial charged particle $\xi $ mass
parameter 
\begin{equation}
m_{phys}:=\ \bar{m}_{es},  \label{3.23d}
\end{equation}%
being exactly equal to the relativistic charged particle $\xi $
electromagnetic mass, as it was assumed by H. Lorentz and Abraham.

To determine the damping radiation force $k(t)\in \mathbb{E}^{3},$ we can
make use of the Lorentz type force expression \ (\ref{3.17}) and obtain,
similarly to \cite{Jack}, up to $O(1/c^{4})$ accuracy, the resulting
self-interecting Abraham-Lorentz type force expression. Thus, owing to the
zero net foirce condition, we have that 
\begin{equation}
d\pi _{p}/dt+F_{s}=0,  \label{3.23e}
\end{equation}%
where the Lorentz force $\ $%
\begin{eqnarray}
\ F_{s} &=&-\ \frac{1}{2c}\int_{\mathbb{\Omega }_{-}(\xi )}d^{3}r\rho (t,r)%
\frac{d}{dt}A(t,r)-\frac{1}{2c}\int_{\mathbb{\Omega }_{+}(\xi )\cup \mathbb{%
\Omega }_{-}(\xi )}d^{3}r\rho (t,r)\frac{d}{dt}A(t,r)-  \label{3.24} \\
&&  \notag \\
&&-\frac{1}{2}\int_{\mathbb{\Omega }_{-}(\xi )}d^{3}r\rho (t,r)\nabla
\varphi (t,r)\ (1-|u/c|^{2})-\frac{1}{2}\int_{\mathbb{\Omega }_{+}(\xi )\cup 
\mathbb{\Omega }_{-}(\xi )}d^{3}r\rho (t,r)\nabla \varphi (t,r)\
(1-|u/c|^{2}).  \notag
\end{eqnarray}%
This expression easily follows from the least action condition $\delta
S^{(t)}=0,\ $\ where $\ S^{(t)}:=\int_{t_{1}}^{t_{2}}\mathcal{L}%
_{f-p}^{(t)}dt$ \ \ with the Lagrangian function given by the derived above
Landau-Lifschitz type expression \ (\ref{3.19}), and the potentials $%
(\varphi ,A)\in T^{\ast }(M^{4})$ given by the Lienard-Wiechert expressions
\ (\ref{3.18}). Followed by calculations similar to those of \cite{Jack,Bart}%
, \ from (\ref{3.24}) and (\ref{3.18}) one can obtain, within the assumed
before uniform shell electron model, for small $|u/c|\ll 1\ $ and slow
enough acceleration \ \ that%
\begin{equation}
\begin{array}{c}
F_{s}=\ \sum_{n\in \mathbb{Z}_{+}}\frac{(-1)^{n+1}}{2n!c^{n}}%
(1-|u/c|^{2})[\int_{\mathbb{\Omega }_{-}(\xi )}\rho (t,r)d^{3}r(\cdot )+ \\ 
\\ 
+\int_{\mathbb{\Omega }_{+}(\xi )\cup \mathbb{\Omega }_{-}(\xi )}\rho
(t,r)d^{3}r(\cdot )]\ \int_{\mathbb{\Omega }_{+}(\xi )}d^{3}r^{\prime }\frac{%
\partial ^{n}}{\partial t^{n}}\rho (t,r^{\prime })\nabla |r-r^{\prime
}|^{n-1\ }+ \\ 
\\ 
+\ \sum_{n\in \mathbb{Z}_{+}}\frac{(-1)^{n+1}}{2n!c^{n+2}}[\int_{\mathbb{%
\Omega }_{-}(\xi )}\rho (t,r)d^{3}r(\cdot )+ \\ 
\\ 
+\int_{\mathbb{\Omega }_{+}(\xi )\cup \mathbb{\Omega }_{-}(\xi )}\rho
(t,r)d^{3}r(\cdot )]\ \int_{\mathbb{\Omega }_{+}(\xi )}d^{3}r^{\prime
}|r-r^{\prime }|^{n-1}\frac{\partial ^{n+1}}{\partial t^{n+1}}J(t,r^{\prime
}]= \\ 
\\ 
=\ \sum_{n\in \mathbb{Z}_{+}}\frac{(-1)^{n+1}}{2n!c^{n+2}}%
(1-|u/c|^{2})[\int_{\mathbb{\Omega }_{-}(\xi )}\rho (t,r)d^{3}r(\cdot )+ \\ 
\\ 
+\int_{\mathbb{\Omega }_{+}(\xi )\cup \mathbb{\Omega }_{-}(\xi )}\rho
(t,r)d^{3}r(\cdot )]\int_{\mathbb{\Omega }_{+}(\xi )}d^{3}r^{\prime }\frac{%
\partial ^{n=2}}{\partial t^{n+2}}\rho (t,r^{\prime })\nabla |r-r^{\prime
}|^{n+1}+ \\ 
\\ 
+\sum_{n\in \mathbb{Z}_{+}}\frac{(-1)^{n+1}}{2n!c^{n+2}}[\int_{\mathbb{%
\Omega }_{-}(\xi )}\rho (t,r)d^{3}r(\cdot )+ \\ 
\\ 
+\int_{\mathbb{\Omega }_{+}(\xi )\cup \mathbb{\Omega }_{-}(\xi )}\rho
(t,r)d^{3}r(\cdot )]\int_{\mathbb{\Omega }_{+}(\xi )}d^{3}r^{\prime
}|r-r^{\prime }|^{n-1}\frac{\partial ^{n+1}}{\partial t^{n+1}}J(t,r^{\prime
}).%
\end{array}
\label{3.24a}
\end{equation}%
The relationship above can be rewritten, owing to the charge continuity
equation \ (\ref{3.1}), giving rise to the radiation force expression 
\begin{equation}
\begin{array}{c}
\\ 
F_{s}=\ \sum_{n\in \mathbb{Z}_{+}}\frac{(-1)^{n\ }}{2n!c^{n+2}}%
(1-|u/c|^{2})[\int_{\mathbb{\Omega }_{-}(\xi )}\rho (t,r)d^{3}r(\cdot
)+\int_{\mathbb{\Omega }_{+}(\xi )\cup \mathbb{\Omega }_{-}(\xi )}\rho
(t,r)d^{3}r(\cdot )]\times \\ 
\\ 
\times \int_{\mathbb{\Omega }_{+}(\xi )}d^{3}r^{\prime }|r-r^{\prime }|^{n-1}%
\frac{\partial ^{n+1}}{\partial t^{n+1}}\left( \frac{\ J(t,r^{\prime })}{n+2}%
+\frac{n-1}{n+2}\frac{<r-r^{\prime },J(t,r^{\prime })>(r-r^{\prime })}{%
|r-r^{\prime }|^{2}}\right) + \\ 
\\ 
+\ \sum_{n\in \mathbb{Z}_{+}}\frac{(-1)^{n+1}}{2n!c^{n+2}}[\int_{\mathbb{%
\Omega }_{-}(\xi )}\rho (t,r)d^{3}r(\cdot )+\int_{\mathbb{\Omega }_{+}(\xi
)\cup \mathbb{\Omega }_{-}(\xi )}\rho (t,r)d^{3}r(\cdot )]\ \int_{\mathbb{%
\Omega }_{+}(\xi )}d^{3}r^{\prime }|r-r^{\prime }|^{n-1}\frac{\partial ^{n+1}%
}{\partial t^{n+1}}J(t,r^{\prime })= \\ 
\\ 
=\ \sum_{n\in \mathbb{Z}_{+}}\frac{(-1)^{n+1}}{2n!c^{n+2}}%
(1-|u/c|^{2})[\int_{\mathbb{\Omega }_{-}(\xi )}\rho (t,r)d^{3}r(\cdot
)+\int_{\mathbb{\Omega }_{+}(\xi )\cup \mathbb{\Omega }_{-}(\xi )}\rho
(t,r)d^{3}r(\cdot )]\ \times \\ 
\\ 
\times \int_{\mathbb{\Omega }_{+}(\xi )}d^{3}r^{\prime }|r-r^{\prime }|^{n-1}%
\frac{\partial ^{n+1}}{\partial t^{n+1}}\left( \frac{\ J(t,r^{\prime })}{n+2}%
+\frac{n-1}{n+2}\frac{|r-r^{\prime },u|^{2}J(t,r^{\prime })}{|r-r^{\prime
}|^{2}|u|^{2}}\right) + \\ 
\\ 
+\ \sum_{n\in \mathbb{Z}_{+}}\frac{(-1)^{n+1}}{2n!c^{n+2}}[\int_{\mathbb{%
\Omega }_{-}(\xi )}\rho (t,r)d^{3}r(\cdot )+\int_{\mathbb{\Omega }_{+}(\xi
)\cup \mathbb{\Omega }_{-}(\xi )}\rho (t,r)d^{3}r(\cdot )]\ \int_{\mathbb{%
\Omega }_{+}(\xi )}d^{3}r^{\prime }|r-r^{\prime }|^{n-1}\frac{\partial ^{n+1}%
}{\partial t^{n+1}}J(t,r^{\prime }). \\ 
\end{array}
\label{3.24b}
\end{equation}%
Now, having applied to \ (\ref{3.24b}) the rotational symmetry property for
calculation of the internal integrals, one easily obtains in the case of a
charged particle $\xi $ uniform shell model that 
\begin{equation*}
\begin{array}{c}
F_{s}=\ \sum_{n\in \mathbb{Z}_{+}}\frac{(-1)^{n}}{2n!c^{n+2}}%
(1-|u/c|^{2})[\int_{\mathbb{\Omega }_{-}(\xi )}\rho (t,r)d^{3}r(\cdot
)+\int_{\mathbb{\Omega }_{+}(\xi )\cup \mathbb{\Omega }_{-}(\xi )}\rho
(t,r)d^{3}r(\cdot )]\ \times \\ 
\\ 
\times \int_{\mathbb{\Omega }_{+}(\xi )}d^{3}r^{\prime }|r-r^{\prime }|^{n-1}%
\frac{\partial ^{n+1}}{\partial t^{n+1}}\left( \frac{\ J(t,r^{\prime })}{n+2}%
+\frac{(n-1)J(t,r^{\prime })}{3(n+2)}\ \right) + \\ 
\\ 
+\ \ \sum_{n\in \mathbb{Z}_{+}}\frac{(-1)^{n+1}}{2n!c^{n}}[\int_{\mathbb{%
\Omega }_{-}(\xi )}\rho (t,r)d^{3}r(\cdot )+\int_{\mathbb{\Omega }_{+}(\xi
)\cup \mathbb{\Omega }_{-}(\xi )}\rho (t,r)d^{3}r(\cdot )]\ \int_{\mathbb{%
\Omega }_{+}(\xi )}d^{3}r^{\prime }\frac{|r-r^{\prime }|^{n+1}}{c^{2}}\frac{%
\partial ^{n+1}}{\partial t^{n+1}}J(t,r^{\prime })\ = \\ 
\end{array}%
\end{equation*}%
\begin{equation}
\begin{array}{c}
=\frac{d}{dt}[\sum_{n\in \mathbb{Z}_{+}}\frac{\ (-1)^{n+1}}{6n!c^{n+2}}%
[\int_{\mathbb{\Omega }_{-}(\xi )}\rho (t,r)d^{3}r(\cdot )+\int_{\mathbb{%
\Omega }_{+}(\xi )\cup \mathbb{\Omega }_{-}(\xi )}\rho (t,r)d^{3}r(\cdot )]\
\times \\ 
\\ 
\times \int_{\mathbb{\Omega }_{+}(\xi )}d^{3}r^{\prime }|r-r^{\prime }|^{n-1}%
\frac{\partial ^{n}}{\partial t^{n}}J(t,r^{\prime })-\sum_{n\in \mathbb{Z}%
_{+}}\frac{(-1)^{n\ }|u|^{2}}{6n!c^{n+4}})[\int_{\mathbb{\Omega }_{-}(\xi
)}\rho (t,r)d^{3}r(\cdot )+ \\ 
\\ 
+\int_{\mathbb{\Omega }_{+}(\xi )\cup \mathbb{\Omega }_{-}(\xi )}\rho
(t,r)d^{3}r(\cdot )]\ \int_{\mathbb{\Omega }_{+}(\xi )}d^{3}r^{\prime
}|r-r^{\prime }|^{n-1}\frac{\partial ^{n}}{\partial t^{n}}J(t,r^{\prime })].%
\end{array}
\label{3.24c}
\end{equation}%
Now, having took into account the integral expressions \ (\ref{3.19}), one
finds from \ (\ref{3.24c}) that up to the $O(1/c^{4})$ accuracy the
following radiation reaction force expression 
\begin{align}
\ F_{s}& =-\frac{d}{dt}\left( \frac{\mathcal{E}_{es}}{c^{2}}u\right) +\frac{d%
}{dt}\left( \frac{\mathcal{E}_{es}}{2c^{2}}\ |u/c|^{2}\ u(t)\right) +\frac{%
2\xi ^{2}}{3c^{3}}\frac{d^{2}u}{dt^{2}}+O(1/c^{4})=  \label{3.24d} \\
&  \notag \\
& =-\frac{d}{dt}\left( m_{es}(1-\frac{|u/c|^{2}}{2})u\right) +\frac{2\xi ^{2}%
}{3c^{3}}\frac{d^{2}u}{dt^{2}}+O(1/c^{4})=  \notag \\
&  \notag \\
& =-\frac{d}{dt}\left( m_{es}(1-|u/c|^{2})^{1/2}u\right) \ \ +\frac{2\xi ^{2}%
}{3c^{3}}\frac{d^{2}u}{dt^{2}}+O(1/c^{4})=  \notag \\
&  \notag \\
& =-\frac{d}{dt}(\bar{m}_{es}u-\frac{2\xi ^{2}}{3c^{3}}\frac{du}{dt}%
)+O(1/c^{4})\   \notag
\end{align}%
holds. \ We mention here that following the reasonings from \cite%
{Beck,Moro,Puth}, in the expressions above there is taken into account an
additional hidden and the velocity $u\in T(\mathbb{R}^{3})$ directed
electrostatic Coulomb surface self-force, acting only on the \textit{front
half part} of the spherical electron shell. As a result, from (\ref{3.23e}),
\ (\ref{3.24}) and the relationship \ (\ref{3.21}) one obtains that the
electron momentum 
\begin{equation}
\pi _{p}:=\bar{m}_{es}u-\frac{2\xi ^{2}}{3c^{3}}\frac{du}{dt}=\bar{m}%
_{es}u-k(t),  \label{3.24e}
\end{equation}%
thereby defyning both the radiation reaction momentum $k(t)=\frac{2\xi ^{2}}{%
3c^{3}}\frac{du}{dt}$ and the corresponding radiation reaction force 
\begin{equation}
F_{r}=\frac{2\xi ^{2}}{3c^{3}}\frac{d^{2}u}{dt^{2}}+O(1/c^{4}),  \label{3.25}
\end{equation}%
coincides exactly with the classical Abraham--Lorentz--Dirac expression.
Moreover, it also follows that the observable physical shell model electron
inertial mass 
\begin{equation}
m_{ph}=\ m_{es}:=\mathcal{E}_{es}/c^{2},  \label{3.25a}
\end{equation}%
$\ $\ being completely of the electromagnetic origin, giving rise to the
final force expression 
\begin{equation}
\frac{d}{dt}(m_{ph}u)=\frac{2\xi ^{2}}{3c^{3}}\frac{d^{2}u}{dt^{2}}%
+O(1/c^{4}).  \label{3.26}
\end{equation}%
This means, in particular, that the real physically observed
\textquotedblleft inertial\textquotedblright\ mass $m_{ph}$\ of an electron
within the uniform shell model is strongly determined by its electromagnetic
self-interaction energy $\mathcal{E}_{es}$. A similar statement there was
recently demonstrated using completely different approaches in \cite%
{Puth,Moro}, based on the vacuum Casimir effect considerations. Moreover,
the assumed above boundedness of the electrostatic self-energy $\mathcal{E}%
_{es}$ appears to be completely equivalent to the existence of so-called
intrinsic Poincar\'{e} type \textquotedblleft \textit{tensions}%
\textquotedblright , analyzed in \cite{Beck,Moro}, and to the existence of a
special compensating Coulomb \textquotedblleft \textit{pressure}%
\textquotedblright , suggested in \cite{Puth}, guaranteeing the observable
electron stability.

\begin{remark}
Some years ago there was suggested in the work \cite{MaPi} a "solution" to
the mentioned before $"4/3$-electron mass" problem, expressed by the
physical mass mass relationship \ (\ref{3.25a}) and formulated more than one
hundred years ago by H. Lorentz and M. Abraham. To the regret, the above
mentioned "solution" appeared to be fake that one can easily observe from
the main not correct assumptions \ on which the work \cite{MaPi} \ has been
based: the first one is about the particle-field momentum conservation,
taken in the form 
\begin{equation}
\frac{d}{dt}(p+\xi A)=0,  \label{3.27}
\end{equation}%
and the second one is a speculation about the $1/2$-coefficient imbedded
into the calculation of the Lorentz type self-interaction force 
\begin{equation}
F:=-\frac{1}{2c}\int_{\mathbb{R}^{3}}d^{3}r\rho (t;r)\partial
A(t;r)/\partial t,  \label{3.28}
\end{equation}%
being not correctly argued by the reasoning that the expression \ (\ref{3.28}%
) represents "... the interaction of a given element of charge with all
other parts, otherwise we count twice that reciprocal action" (cited from \ 
\cite{MaPi}, page 2710). This claim is fake as there was not taken into
account the important fact that the interaction in the integral \ (\ref{3.28}%
) is, in reality, retarded and its impact into it should be considered as
that calculated for two virtually different charged particles, as it has
been done in the classical works of H. Lorentz and M. Abraham. Subject to
the first assumption \ (\ref{3.27}) it is enough to recall that a vector of
the field momentum $\xi A\in \mathbb{E}^{3}$ is not independent and is,
within the charged particle model considered, strongly related with the
local flow of the electromagnetic potential energy in the Lorentz constraint
form:%
\begin{equation}
\partial \varphi /\partial t+<\nabla ,A\ >=0,  \label{3.29}
\end{equation}%
under which there hold \ the exploited in the work \cite{MaPi} the
Lienard-Wiechert expressions \ (\ref{3.17}) potentials\ for calculation of
the integral \ (\ref{3.28}). Thus, the equation \ (\ref{3.27}), following
the classical Newton second law, should be replaced by%
\begin{equation}
\frac{d}{dt^{\prime }}(p^{\prime }+\xi A^{\prime })=-\nabla (\xi \varphi
^{\prime }),  \label{3.30}
\end{equation}%
written with respect to the reference frame $\mathcal{K}(t^{\prime };r%
\mathcal{)}$ subject to which the charged particle $\xi \ $ is at rest.
Taking into account that with respect to the laboratory reference frame $%
\mathcal{K}_{t}$ there hold the relativistic relationships $dt=dt^{\prime
}(1-|u|^{2}/c^{2})^{1/2}\ $\ and $\varphi ^{\prime }=\varphi
(1-|u|^{2}/c^{2})^{1/2},$ \ \ from (\ref{3.30}) one easily obtains that 
\begin{equation}
\begin{array}{c}
\frac{d}{dt}(p+\xi A)=-\xi \nabla \varphi (1-|u|^{2}/c^{2})= \\ 
\\ 
=-\xi \nabla \varphi +\frac{\xi }{c}\nabla <u,u\varphi /c>=-\xi \nabla
\varphi +\frac{\xi }{c}\nabla <u,A>.%
\end{array}
\label{3.31}
\end{equation}%
Here we made use of the well-known relationship $A=u\varphi /c$ for the
vector potential generated by this charged particle $\xi $ moving in space
with the velocity $u\in T(\mathbb{R}^{3})$ with respect to the laboratory
reference frame $\mathcal{K}_{t}.$ Based now on the equation \ (\ref{3.31})
one can derive the final expression for the evolution of the charged
particle $\xi $ momentum:%
\begin{eqnarray}
dp/dt &=&-\xi \nabla \varphi -\frac{\xi }{c}dA/dt+\frac{\xi }{c}\nabla <u,A>=
\label{3.32} \\
&&  \notag \\
&=&-\xi \nabla \varphi -\frac{\xi }{c}\partial A/\partial t-\frac{\xi }{c}%
<u,\nabla >A+\frac{\xi }{c}\nabla <u,A>=  \notag \\
&&  \notag \\
&=&\xi E+\frac{\xi }{c}u\times (\nabla \times A)=\xi E+\frac{\xi }{c}u\times
B,  \notag
\end{eqnarray}%
that is exactly the well known Lorentz force expression, used in the works
of H. Lorentz and M. Abraham.
\end{remark}

Recently enough there appeared other interesting works devoted to this "$4/3$%
-electron mass" problem, amongst which we would like to mention \cite%
{Moro,Puth}, whose argumentations are close to each other and based on the
charged shell electron model, within which there is assumed a virtual
interaction of the electron with the ambient "dark" radiation energy. The
latter was first clearly demonstrated in \cite{Puth}, \ \ where a suitable
compensation mechanism of the related singular electrostatic Coulomb
electron energy and the wide band vacuum electromagnetic radiation energy
fluctuations deficit inside the electron shell was shown to be harmonically
realized as the electron shell radius $a\rightarrow 0.$ Moreover, this
compensation happens exactly when the induced outward directed electrostatic
Coulomb pressure on the whole electron coincides, up to the sign, with that
induced by the mentioned above vacuum electromagnetic energy fluctuations
outside the electron shell, since there was manifested their absence inside
the electron shell.

Really, the outward directed electrostatic spatial Coulomb pressure on the
electron equals 
\begin{equation}
\eta _{coul}:=\left. \lim_{a\rightarrow 0}\frac{\varepsilon _{0}|E|^{2}}{2}%
\right\vert _{r=a}=\lim_{a\rightarrow 0}\frac{\xi ^{2}}{32\varepsilon
_{0}\pi ^{2}a^{4}},  \label{3.33}
\end{equation}%
where $E=\frac{\xi r}{4\pi \varepsilon _{0}|r|^{3}}\in \mathbb{E}^{3}$ is
the electrostatic field at point $r\in \mathbb{R}\ $\ subject to the
electron center at the point $r=0\in \mathbb{R}.$ The related inward
directed vacuum electromagnetic fluctuations spatial pressure equals 
\begin{equation}
\eta _{vac}:=\lim_{\Omega \rightarrow \infty }\frac{1}{3}\int_{0}^{\Omega }d%
\mathcal{E}(\omega ),  \label{3.34}
\end{equation}%
where $d\mathcal{E}(\omega )$ is the electromagnetic energy fluctuations
density for a frequency $\omega \in \mathbb{R},$ and $\Omega \in \mathbb{R}$
is the corresponding electromagnetic frequency cutoff. The integral \ (\ref%
{3.34}) can be calculated if to take into account the quantum statistical
recipe \cite{Feyn,Huan,BoBo} that 
\begin{equation}
d\mathcal{E}(\omega ):=\hbar \omega \frac{d^{3}p(\omega )}{h^{3}},
\label{3.35}
\end{equation}%
where the Plank constant $h:=2\pi \hbar $ and the electromagnetic wave
momentum $p(\omega )\in \mathbb{E}^{3}$ satisfies the relativistic
relationship 
\begin{equation}
|p(\omega )|=\hbar \omega /c.  \label{3.36}
\end{equation}%
Whence by substituting \ (\ref{3.36}) into \ (\ref{3.35}) one obtains 
\begin{equation}
d\mathcal{E}(\omega )=\frac{\hbar \omega ^{3}}{2\pi ^{2}c^{3}}d\omega ,
\label{3.37}
\end{equation}%
which entails, owing to \ (\ref{3.34}), the following vacuum electromagnetic
energy fluctuations spatial pressure%
\begin{equation}
\eta _{vac}=\lim_{\Omega \rightarrow \infty }\frac{\hbar \Omega ^{4}}{24\pi
^{2}c^{3}}.  \label{3.38}
\end{equation}

For the charged electron shell model to be stable at rest it is necessary to
equate the inward \ (\ref{3.38}) and outward \ (\ref{3.33}) spatial
pressures:%
\begin{equation}
\lim_{\Omega \rightarrow \infty }\frac{\hbar \Omega ^{4}}{24\pi ^{2}c^{3}}%
=\lim_{a\rightarrow 0}\frac{\xi ^{2}}{32\varepsilon _{0}\pi ^{2}a^{4}},
\label{3.39}
\end{equation}%
giving rise to the balance electron shell radius $a_{b}\rightarrow 0$
limiting condition:%
\begin{equation}
a_{b}=\lim_{\Omega \rightarrow \infty }\ \left[ \Omega ^{-1}\left( \frac{%
3\xi ^{2}c^{2}}{2\hbar }\right) ^{1/4}\right] .  \label{3.40}
\end{equation}

Simultaneously we can calculate the corresponding Coulomb and
electromagnetic fluctuations energy deficit inside the electron shell:%
\begin{equation}
\Delta W_{b}:=\frac{1}{2}\int_{a_{b}}^{\infty }\varepsilon
_{0}|E|^{2}d^{3}r-\int_{0}^{a_{b}}d^{3}r\int_{0}^{\Omega }d\mathcal{E}%
(\omega )=\frac{\xi ^{2}}{8\pi \varepsilon _{0}a_{b}}-\frac{\hbar \Omega
^{4}a_{b}^{3}}{6\pi c^{3}}=0,\   \label{3.41}
\end{equation}%
additionally ensuring the electron shell model stability.

Another important consequence from this pressure-energy compensation
mechanism can be derived concerning the electron ienrtial mass $m_{ph}\in 
\mathbb{R}_{+},$ entering the momentum expression \ (\ref{3.24e}) in the
case of the electron slow enough movement. Namely, following the reasonings
from \cite{Moro}, one can observe that during the electton movement there
arises an additional hidden not compensated and velocity $u\in T(\mathbb{R}%
^{3})$ directed electrostatic Coulomb surface self-pressure acting only on
the \textit{front half part} of the electron shell and equal to 
\begin{equation}
\eta _{surf}:=\frac{|E\xi |}{4\pi a_{b}^{2}}\frac{1}{2}=\frac{\xi ^{2}}{%
32\pi \varepsilon _{0}a_{b}^{4}},  \label{3.42}
\end{equation}%
coinciding, evidently, with the already compensated outward directed
electrostatic Coulomb spatial pressure \ (\ref{3.33}). As, evidently, during
the electron motion in space its surface electric current energy flow is not
vanishing \cite{Moro}, one ensues that the electron momentum gains an
additional mechanical impact, which can be expressed as 
\begin{equation}
\pi _{\xi }:=-\eta _{surf}\frac{4\pi a_{b}^{3}}{3c^{2}}u=-\frac{1}{3}\frac{%
\xi ^{2}}{8\pi \varepsilon _{0}a_{b}c^{2}}u=-\frac{1}{3}\bar{m}_{es}u,
\label{3.43}
\end{equation}%
where we took into account that within this electron shell model the
corresponding electrostatic electron mass equals its electrostatic energy 
\begin{equation}
\bar{m}_{es}=\frac{\xi ^{2}}{8\pi \varepsilon _{0}a_{b}c^{2}}.  \label{3.44}
\end{equation}

Thus, one can claim that, owing to the structural stability of the electron
shell model, its generalized self-interaction momentum $\pi _{p}\in T^{\ast
}(\mathbb{R}^{3})$ \ gains during the movement with velocity $u=dr/dt\in T(%
\mathbb{R}^{3})$ the additional backward directed hidden impact \ (\ref{3.43}%
), which can be identified with the back-directed momentum component 
\begin{equation}
\pi _{\xi }=-\frac{1}{3}\bar{m}_{es}u,  \label{3.45}
\end{equation}%
complementing the classical \cite{Jack,Bart} momentum expression $\ $%
\begin{equation}
\pi _{p}=\frac{4}{3}\bar{m}_{es}u,  \label{3.46}
\end{equation}
which can be easily obtained from the Lagrangian expression expression, if
one not to take into account the shading property of the moving uniform
shell electron model. \ Then, owing to \ the additional momentum (\ref{3.45}%
), the full momentum becomes as 
\begin{equation}
\pi _{p}=\pi _{\xi }+\frac{4}{3}\bar{m}_{es}u\ =(-\frac{1}{3}\bar{m}_{es}+%
\frac{4}{3}\bar{m}_{es})u=\bar{m}_{es}u,  \label{3.47}
\end{equation}%
coinciding with that of \ (\ref{3.21}) modulo the radiation reaction\
momentum $k(t)=\frac{2\xi ^{2}}{3c^{3}}\frac{du}{dt},$ \ strongly supporting
the electromagnetic energy origin of the electron inertion mass for the
first time conceived by H. Lorentz and M. Abraham.

\section{Comments}

The electromagnetic mass origin problem was reanalyzed in details within the
Feynman proper time paradigm and related vacuum field theory approach by
means of the fundamental least action principle and the Lagrangian and
Hamiltonian formalisms. The resulting electron inertia appeared to coincide
in part, in the quasi-relativistic limit, with the momentum expression
obtained more than one hundred years ago by M. Abraham and H. Lorentz \cite%
{Abra,Lore-1,Lore-2,Lore-3}, yet it proved to contain an additional hidden
impact owing to the imposed electron stability constraint, which was taken
into account in the original action functional as some preliminarily
undetermined constant component. As it was demonstrated in \cite{Puth,Moro},
this stability constraint can be successfully realized within the charged
shell model of electron at rest, if to take into account the existing
ambient electromagnetic \textquotedblleft dark\textquotedblright\ energy
fluctuations, whose inward directed spatial pressure on the electron shell
is compensated by the related outward directed electrostatic Coulomb spatial
pressure as the electron shell radius satisfies some limiting compatibility
condition. The latter also allows to compensate simultaneously the
corresponding electromagnetic energy fluctuations deficit \ inside the
electron shell, thereby forbidding the external energy to flow into the
electron. In contrary to the lack of energy flow inside the electron shell,
during the electron movement the corresponding internal momentum flow is not
vanishing owing to the nonvanishing hidden electron momentum flow caused by
the surface pressure flow and compensated by the suitably generated surface
electric current flow. As it was shown, \ this backward directed hidden
momentum flow makes it possible to justify the corresponding
self-interaction electron mass expression and to state, within the electron
shell model, the fully electromagnetic electron mass origin, as it has been
conceived by H. Lorentz and M. Abraham and strongly supported by R. Feynman
in his Lectures \cite{Feyn-1}. This consequence is also independently
supported by means of the least action approach, based on the Feynman proper
time paradigm and the suitably calculated regularized retarded \ electric
potential impact into the charged particle Lagrangian function.

The charged particle radiation problem, revisited in this Section, allowed
to conceive the explanation of the charged particle mass as that of a
compact and stable object which should be exerted by a vacuum field
self-interaction energy. The latter can be satisfied iff the expressions (%
\ref{3.19}) hold, thereby imposing on the intrinsic charged particle
structure \cite{Medi} some nontrivial geometrical constraints. Moreover, as
follows from the physically observed particle mass expressions (\ref{3.25a}%
), the electrostatic potential energy being of the \ \ \ self-interaction
origin, contributes into the inertial mass as its main relativistic mass
component.

There exist different relativistic generalizations of the force expression \
(\ref{3.26}), which \ suffer the common physical inconsistency related to
the no radiation effect of \ a charged particle in uniform motion.

Another deeply related \ problem to the radiation reaction force analyzed
above is the search for an explanation to the Wheeler and Feynman reaction
radiation mechanism, called the absorption radiation theory, strongly based
on the Mach type interaction of a charged particle with the ambient vacuum
electromagnetic medium. Concerning this problem, one can also observe some
of its relationships with the one devised here within the vacuum field
theory approach, but this question needs a more detailed and extended
analysis.

\section{\protect\bigskip Acknowledgements}

A.P. would like to convey his cordial thanks to Prof. Hal Puthoff (Institute
for Advanced Studies at Austin, Texas USA) for sending me his original
works, \ and to Jerrold Zacharias Professor of Physics Roman Jackiw
(Department of Physics at the Massachusett Institute of Technology, MT, USA)
for instrumental discussion during his collaborative research stay at the
NJIT, NJ USA during May 20-31, 2015, as well as for the related comments and
useful remarks. The authors also very acknowledged \ to Prof. Denis
Blackmore (NJIT, Newark NJ, \ USA) and Prof. Edward Kapuscik (Institute for
Nuclear Physics at PAS, Krak\'{o}w, Poland) for friendly cooperation and
important discussions. The work of N.B. was supported by the RSF under a
grant 14-50-00005.

\end{document}